%% file: main.tex
\documentclass[conference]{IEEEtran}
\IEEEoverridecommandlockouts
\usepackage{cite}
\usepackage{amsmath,amssymb,amsfonts}
\usepackage{graphicx}
\usepackage{textcomp}
\usepackage{xcolor}

\usepackage[normalem]{ulem}

\newcommand\blfootnote[1]{%
  \begingroup
  \renewcommand\thefootnote{}%
  \footnote{#1}%
  \addtocounter{footnote}{-1}%
  \endgroup
}

\usepackage{progressbar}
\usepackage[linesnumbered,ruled,vlined]{algorithm2e}
\usepackage{amssymb}
\usepackage[colorlinks=true, allcolors=blue]{hyperref}
\usepackage{dirtytalk}
\usepackage{subcaption}
\usepackage{algpseudocode}
\usepackage{indentfirst}
\usepackage[abbreviations]{foreign}  

\usepackage{pdfpages}

\newcommand{\brahms}{\textsc{Brahms}\xspace}
\newcommand{\AP}{\textsc{Aupe}\xspace}
\newcommand{\raptee}{\textsc{Raptee}\xspace}
\newcommand{\basalt}{\textsc{Basalt}\xspace}
\newcommand{\secureCyclon}{\textsc{Secure Cyclon}\xspace}

\begin{document}

\title{\AP: Collaborative Byzantine fault-tolerant peer-sampling
}

\author{}

\author{\IEEEauthorblockN{Augusta Mukam, Joachim Bruneau-Queyreix, Laurent Réveillère}
	\IEEEauthorblockA{
		\textit{University of Bordeaux, Bordeaux-INP, CNRS-LaBRI}\\
		F-33400 Talence, France}
}


\newcommand{\red}[1]{\textcolor{red}{#1}} 
\newcommand{\bluestrike}[1]{\begingroup\color{blue}\sout{#1}\endgroup}

\renewcommand{\bluestrike}[1]{}
\renewcommand{\red}[1]{#1}

\newcommand{\annote}[3]{{
			\colorbox{#3}{\bfseries\sffamily\footnotesize\textcolor{white}{#2}}
			\color{#3}		$\blacktriangleright$\textit{#1}$\blacktriangleleft$}
}

\newcommand{\jbq}[1]{\annote{#1}{JBQ}{blue}}
\newcommand{\am}[1]{\annote{#1}{AM}{violet}}
\newcommand{\lr}[1]{\annote{#1}{LR}{red}}

\newcommand{\eval}[1]{
	\progressbar[linecolor=blue, filledcolor=green]{\fpeval{#1/100}}\llap{\raisebox{1.5pt}{\tiny$ \fpeval{#1}\%$}\hspace{-.6cm}}
}
\maketitle

\begin{abstract}
	Peer sampling is a crucial primitive in distributed systems, used to manage overlays and disseminate information in large-scale scenarios such as permissionless blockchain systems.
	Its purpose is to maintain and regularly update a local and partial snapshot, or view, of the complete system's membership.
	These protocols are often targeted by malicious actors who aim to disrupt higher-level protocols.
	Typically, an adversary who controls a set of Byzantine nodes attempts to manipulate how legitimate nodes perceive the presence of Byzantine ones by increasing their representation in the view of honest nodes.
	While state-of-the-art Byzantine-tolerant peer sampling protocols mitigate this bias, their effectiveness decreases significantly as the number of malicious nodes increases.
	This paper introduces \AP, the first collaborative Byzantine-tolerant peer sampling protocol that leverages the presence of trusted nodes, such as Intel's SGX capable devices, to collaboratively track the spread of identifiers in the system and locally debias the representation of Byzantine nodes.
	Simulations with 10,000 nodes demonstrate that \AP outperforms state-of-the-art solutions, achieving near-perfect resilience even when faced with an adversary controlling 26\% of the nodes.
	Overall, by including as few as 10\% of trusted nodes, \AP increases the \bluestrike{resilience}\red{tolerance} of ~\brahms by up to 60\% while limiting the impact of the adversary's attack, even when possessing up to 40\% of the nodes.

\end{abstract}

\blfootnote{The research leading to these results has received funding from the French National Research Agency (ANR) under grant ANR-21-CE25-0021-03}

\begin{IEEEkeywords}
	Gossip, Peer Sampling, Distributed System,
	Byzantine tolerance, Eclipse Attacks
\end{IEEEkeywords}

\input{sections/intro.tex}
\input{sections/system_model.tex}
\input{sections/protocol.tex}
\input{sections/eval.tex}

\input{sections/sota.tex}

\input{sections/conclusion.tex}

\bibliographystyle{IEEEtran}
\bibliography{sample.bib}

\end{document}

%% file: sections/intro.tex
\section{Introduction}


Gossip-based peer sampling is a well-known paradigm used for building large-scale and dynamic applications, primarily for information dissemination, overlay management, and recently in blockchain consensus protocols~\cite{nakamoto2008bitcoin, yahyaoui2024tolerating, Matos:2013,korkmaz2022depth, Voulgaris:2005, Heilman:2015, JESI20102086}.
For instance, they are currently largely employed in public blockchain systems with several thousands of nodes.
These applications depend heavily on the system’s ability to manage and maintain up-to-date information about its members.
Gossip-based protocols have been designed to achieve desirable properties~\cite{Jelasity:2007} such as scalability, robustness to node failures, and a low probability of partitions even when nodes join or leave.

In this context, the role of peer sampling services becomes crucial.
These services aim to maintain knowledge of active nodes in the system by continuously providing uniformly and randomly selected samples of node descriptors from the entire population.
Typically, each node has a local knowledge of the system called a \emph{view} which consists of a set of neighbor identifiers or descriptors.
This view is periodically updated by exchanging messages with some of its neighbors.
Typically, during a \emph{round} of the peer sampling protocol execution, each node selects at least one peer from its view to push its own identifier or exchange known identifiers.
This process spreads knowledge of active nodes throughout the system, ensuring global connectivity.

Peer sampling services face challenges in achieving their goal of randomly selecting partners.
Malicious, \ie, Byzantine, nodes can manipulate the peer sampling protocol to promote specific nodes within their member group as honest nodes.
These nodes engage in such behavior to increase their representation in the views of correct nodes.
As the percentage of malicious descriptors in the views of honest nodes increases, the probability of selecting these malicious identifiers for request exchanges also increases.
Consequently, the views of honest nodes gradually become saturated with malicious descriptors.
An attacker controlling multiple nodes can exploit this vulnerability to poison the views of honest nodes.
This can lead to scenarios where the views of some honest nodes consist entirely of these malicious identifiers.
As a result, the poisoned nodes may become isolated from the system, unable to connect with other honest nodes.
Additionally, malicious nodes can also gain a leading position within the system, by becoming over-represented in the views of honest nodes, enabling them to launch network attacks such as DoS or target higher-level protocols like consensus.
A notable example of this vulnerability is the Eclipse attack in Bitcoin's peer sampling protocol~\cite{Heilman:2015}.
In this attack, Byzantine nodes exclude honest nodes and control all their connections.
This manipulation enables attackers to present a corrupted version of the blockchain and even manipulate tokens of victim nodes.

Two distinct approaches have been explored to mitigate the presence of Byzantine behaviors in peer-sampling protocols.
Protocols such as SPS~\cite{JESI20102086} and \secureCyclon~\cite{Antonov:2023} employ fault-detection to identify malicious nodes and produce proofs of misbehavior to blame and evict the latter nodes from the system.
Other protocols, such as \brahms~\cite{Bortnikov:2008}, along with its extensions \raptee~\cite{Pigaglio:2022} or \basalt~\cite{basalt:2023}, aim to tolerate Byzantine nodes by masking their behaviors.
These protocols have demonstrated their ability to tolerate a limited number of malicious nodes in the system, typically never larger than 30\% of the system's population.
For example, in \brahms, the views of honest nodes contain 71\% and 80\% of Byzantine identifiers as 24\% and 28\% of the nodes in the system are malicious, respectively.
\raptee improves the resilience of \brahms by 21\% by leveraging the presence of 10\% of trusted nodes operating on trusted execution environments (TEE), \eg Intel's SGX~\cite{sgx}.
\basalt outperforms \brahms in resisting Byzantine attacks; however, its effectiveness declines once the proportion of Byzantine nodes exceeds 20\%.

We introduce \AP, yet another extension of \brahms designed to withstand Byzantine attacks even when a significant number of nodes are compromised.
The \AP protocol enhances Byzantine resistance by equipping nodes with a \emph{Set Cleaner} that consists of two key components: a tracking component that records the frequency of received identifiers, and a debiasing component responsible for transforming a given set of IDs into a set that more closely matches the expected uniform distribution of the IDs observed thus far.
This reduces the likelihood of frequently encountered nodes being resampled.
Similarly to \raptee, \AP also benefit from the presence of nodes running on a trusted execution environment.
In this paper, we consider Intel SGX, but any TEE technology providing similar code integrity and remote-attestation properties would be a fit for
our approach, \eg ARM’s TrustZone~\cite{trustZone}.
In \AP, trusted nodes track collaboratively and in a distributed fashion the dissemination of identifiers in the system by merging their respective tracking component.
This merge strategy allows them to quickly gain a better understanding of the frequencies of node ID advertisement in the overall system and provides a larger basis of information to operate the debiasing mechanism.
Hence, trusted nodes collectively serve as a source of less biased views for the rest of the system.
Our simulation results including 10,000 nodes show that \AP achieves near-perfect resilience with 26\% of malicious nodes, where the average percentage of Byzantine samples in honest nodes' views converges to the proportion of malicious nodes in the system.

In section~\ref{sec:sysmodel}, we describe the basis of this work, \brahms, our system model and objectives as well as the attack model of our adversary.
In Section~\ref{sec:ap}, we describe \AP and detail how it complements \brahms with its debiasing approach and
Section~\ref{sec:evaluation} presents the experimental methodology used to evaluate the performance and resilience of \AP under different configurations and discusses the results.
We review the state of the art in Section~\ref{sec:related} before concluding in Section~\ref{sec:conclusion}.


%% file: sections/system_model.tex
\section{Background, System model and Objectives}
\label{sec:sysmodel}


\subsection{\brahms}
\label{s:brahms}
\brahms~\cite{Bortnikov:2008}, is designed for large and dynamic systems prone to Byzantine failures and sybil attacks.
With high probability, \brahms prevents an attacker from creating a partition between correct nodes, and allows each node's view to converge to a uniform random draw of all participating nodes over time.
The core ideas of \brahms lies in its two components used by all nodes: a push-pull gossip-based communication pattern, and a uniform sampler based on min-wise independent permutations~\cite{broder1998min}.
Through the gossip component, each node spreads locally known IDs across the system and maintains a dynamic view $V$ of $l_1$ entries.
With the sampling component, each node maintains a \emph{sample list} $S$ of $l_2$ entries uniformly sampled from the received IDs. 

\begin{figure}[!htb]
  \centering
  \includegraphics[width=.75\linewidth]{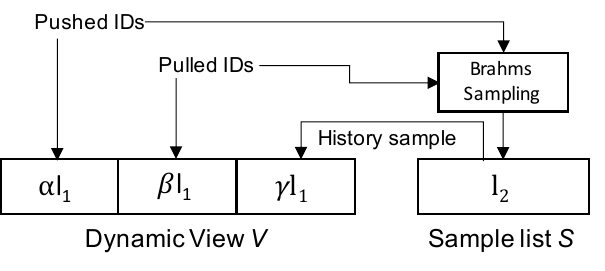}
  \caption{\brahms view computation}
  \label{fig:brahms_view}
\end{figure}

Initially, each node has a list that includes node IDs and addresses obtained from a bootstrap node.
Periodically, through the gossip component, each node selects $\alpha \times l_1$ nodes from its dynamic view $V$ to send push messages containing its own ID.
It also selects $\beta \times l_1$ nodes to send pull-request messages in order to retrieve their views.
At the end of each communication round, the sample list $S$ is updated via the sampling component, while the dynamic view $V$ is renewed by selecting uniformly at random: (i) $\alpha \times l_1$ IDs received from push messages, (ii) $\beta \times l_1$ IDs from pull answers, and (iii) $\gamma \times l_1$ IDs from the sample list (referred to as the \emph{history sample}), with $\alpha+\beta+\gamma = 1$, as shown in Figure~\ref{fig:brahms_view}.

\begin{figure}[!htb]
  \centering
  \includegraphics[width=.85\linewidth]{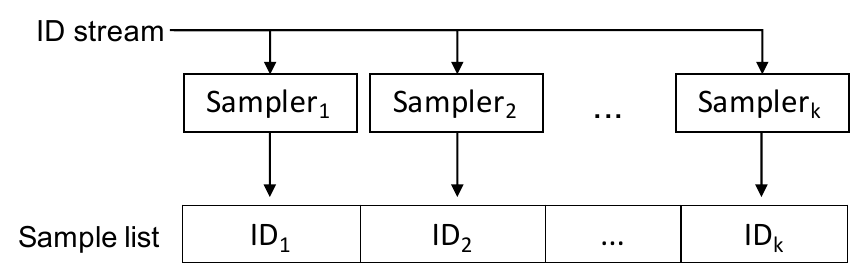}
  \caption{\brahms sampling component}
  \label{fig:brahmssampler}
\end{figure}

Depicted in figure~\ref{fig:brahmssampler}, the sampling component outputs a list of $l_2$ IDs from a stream of identifiers obtained via push and pull messages.
The component is composed of a set of $l_2$ samplers, each of which consists of a hash function with a seed selected pseudo-randomly at bootstrap and a locally stored ID.
For each input ID, a sampler stores and outputs the ID that results in the smallest value from its hash function between the current ID and the stored one.


The four defense mechanisms that prevent partitioning and allow convergence to uniform sampling are the following:
(i)~limited pushes,
(ii)~attack detection and blocking,
(iii)~controlling the contribution of pulls versus pushes, and
(iv)~history sampling.
We detail these mechanisms in the following.
(i)~An adversary could forge identities or flood the system with push requests, leaving correct IDs propagated mainly through pulls and diminishing their representation exponentially.
\brahms assumes a mechanism that limits the message sending rate of nodes, for example, via computational challenges like Merkle's puzzles, virtual currency, \etc
(ii)~Limiting push messages prevents a simultaneous attack on all correct nodes but does not protect against flooding a targeted node.
To do so, \brahms blocks dynamic view updates if more than the expected $\alpha \times l_1$ pushes are received.
This policy slows down progress but its expected impact in the absence of attacks is bounded, and thanks to limited pushes, some nodes make progress even under attack.
(iii)~Node views are threatened by pulls from neighbors more than by adversarial pushes.
Pushes from correct nodes are correct, but pull answers from correct nodes may contain some identifiers of Byzantine nodes.
Hence, the contribution of pushes and pulls to $V$ must be balanced.
\brahms updates $V$ with randomly chosen $\alpha \times l_1$ pushed IDs to protect targeted nodes, and $\beta \times l_1$ pulled IDs to protect the rest.
(iv)~The attack detection and blocking technique slows down a targeted attack but cannot prevent it completely.
A node victim of targeted pushes will pull more IDs from Byzantine nodes, send fewer pushes to correct ones, causing its system-wide representation to decrease, hence receiving fewer correct pushes.
\brahms overcomes such an attack using a self-healing mechanism by incorporating in the view an unbiased historical sample of $\gamma \times l_1$ IDs from the sample list $S$.
Once some correct ID becomes the permanent sample of the node under attack (or if the node’s ID becomes a permanent sample of another correct node) the threat of isolation is eliminated.

\subsection{System model}

We consider a system of $N$ active nodes, each identified by a unique identifier ID and that communicates by exchanging messages on a routed network.
This system runs a gossip-based peer sampling protocol for peer discovery that operates periodically in rounds.
We assume that the system is at a time $T_0$ where  no nodes join or leave the system.
Each node possesses a local view of $l_1$ entries representing its neighbors.

The system is composed of a fraction $f < 1$ of malicious nodes managed by an attacker trying to over-represent itself, a fraction $t$ of nodes operating on a trusted execution environment, referred to as \emph{trusted nodes}, and a fraction $h=1-f-t$ of honest nodes following the instructions of the peer sampling protocol \AP.

Honest nodes execute \AP, the extension of the iconic \brahms peer-sampling protocol. 
\AP relies on both the push-pull gossip and sampling components of \brahms as well as on its defense mechanisms.
Push messages include only the sender’s ID, while responses to pull requests contain the full view of the requested node.
\emph{\AP nodes} integrate a local component, coined \emph{\AP's set cleaner}, responsible for correcting the representation bias in favor of Byzantine nodes that exist in the sets of received IDs at the end of each round.
Similarly, trusted nodes adopt the same behavior as honest nodes.
In addition, interactions among trusted nodes are designed to boost their set cleaner and help them better debias their sets of received IDs, thus acting as source of slightly less biased membership information for honest nodes.

\subsection{Attack model and design goal}
\label{ss:attack}


We consider an adversary controlling all Byzantine nodes, with the goal of undermining the system by manipulating and increasing honest nodes' perception of the proportion of Byzantine nodes.
The adversary has access to the system’s global membership, including Byzantine and correct nodes, but is unaware of the location or amount of trusted nodes.
In the original \brahms paper~\cite{Bortnikov:2008}, the authors prove that a balanced attack, which spreads faulty pushes evenly among correct nodes, maximizes the expected system-wide fraction of faulty IDs.
Additionally, thanks to its sampling mechanism, \brahms sustains targeted attacks where the adversary tries to partition the network by targeting a subset of nodes and sending more pushes than in balanced attacks.
As \AP is built on top of \brahms and inherits its properties, our adversary exclusively advertises Byzantine IDs to honest ones via answers to pull requests and evenly balanced push messages.


We rule out Sybil attacks~\cite{douceurSybil2002} by relying on the Sybil resilience of \brahms that limits the message sending rate of nodes via computational challenges like Merkle’s puzzles, virtual currency, \etc
We assume that trusted nodes can only crash fault.
They cannot act maliciously even though Byzantine IDs can bias their view.
Trusted nodes operate on Intel's SGX framework, and we trust Intel for the certification of genuine SGX-enabled CPUs, we also assume that the code running inside enclaves is properly attested before being provided with secrets.
We assume Byzantine nodes can neither break cryptographic primitives nor read data available in the trusted environment of trusted nodes.
We also assume that it is impractical for an adversary to extract secrets from enclaves using side-channel attacks, having no physical access or means to colocate a process on a target machine.
Communications between any two nodes, including trusted ones, are ciphered with symmetric encryption to protect against an eavesdropping adversary.
\bluestrike{We}\red{Finally, we} consider that the adversary does not have global access to communication links in the network that do not involve itself and cannot, therefore, infer information from communication patterns.




%% file: sections/protocol.tex
\section{The \AP protocol}\label{sec:ap}



 \subsection{\AP set cleaner}
\label{ss:dm}

\begin{figure}[!t]
     \centering
     \includegraphics[width=0.9\linewidth]{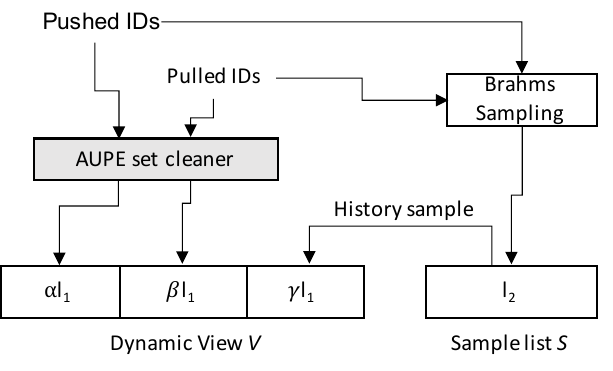}
     \caption{Overview of \AP's view computation}
     \label{fig:AP_view}
 \end{figure}

In \brahms, the view of honest nodes is composed of $\gamma *l_1$ identifiers from \brahms's history sample. 
This provides Byzantine resilience once enough node identifiers have passed through the sampling component, but only for these $\gamma *l_1$ entries.
The remaining $(\alpha+\beta)l_1$ entries are selected directly from the sets of received identifiers that are biased towards the representation for malicious node IDs due to the attack led by Byzantine nodes.
To mitigate this problem, \AP integrates a \emph{Set Cleaner} that is responsible for mitigating this bias during the computation of each node's view, as depicted in Figure~\ref{fig:AP_view}, 

 \begin{figure}[!t]
     \centering
     \includegraphics[width=0.97\linewidth]{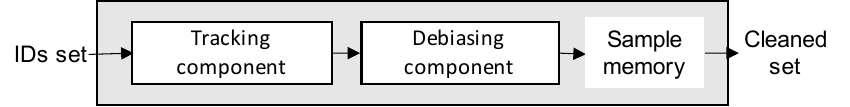}
     \caption{Overview of \AP's set cleaner}
     \label{fig:overview}
 \end{figure}

At the end of each round, the sets of received identifiers from push and pull requests are injected into the Set Cleaner of \AP. 
The Set Cleaner is composed of two components: a tracking component that records the occurrences of received IDs from the beginning of the protocol's execution, and a debiasing component responsible for transforming a given ID set into a set that better matches the expected uniform distribution of the IDs seen up to that point.
The output set is used to select the required number of identifiers to be used for the view construction of \AP.
Figure~\ref{fig:overview} depicts the general operations of \AP's Set Cleaner.
In \AP's case, the two sets resulting from the pulled and pushed IDs are passed through the Set Cleaner, and two sets of $\alpha*l_1$ and $\beta*l_1$ are created to fill the \emph{push} and \emph{pull} parts of \AP's dynamic view.

By reducing the bias introduced by the Byzantine nodes in its sets, the \AP node ends up with less biased views.
This, in turn, helps to spread less biased information about the membership of the system, benefiting other honest nodes.


\textbf{Tracking component}
Each node has a \emph{tracking datastructure} providing the occurrences of received IDs.
This can simply consists of a key-value store where the keys are the received identifiers and the values of their occurrences (i.e., the number of times they were received). 
Each pushed or pulled ID feeds this occurrence table throughout the execution of the protocol.
In systems where the memory available on each node is constrained, \AP \bluestrike{uses} \red{can use} count-min-sketches, \red{which are} fixed-size probabilistic data-structures, to drastically reduce memory footprint.
This comes at the cost of a reduced precision on the identifiers estimated occurrences depending on the actual size of the latter data structure.
Our current implementation relies on the count-min-sketches described by Anceaume et al.~\cite{Anceaume:2013}, where the precision reduction is determined and quantified. 

\textbf{Debiasing component}
Debiasing is performed at the end of each round of the protocol. 
The two sets resulting from the received pulled and pushed IDs are transformed using this component.
Its algorithm is detailed in Algorithm~\ref{alg:Occstart} when considering a tracking component in the form of a key-value store $\Phi$. 
We refer to the approach detailed by Anceaume et al.~\cite{Anceaume:2013} when relying on count-min-sketches.
We consider a \AP node that receives a set $\sigma$ of IDs at the end of round $r$.
For each ID in $\sigma$, the debiasing component produces an ID to build the output set $\sigma^{'}$.
The node maintains a local \emph{sample memory} $\Gamma$ of size \bluestrike{$sm$.}\red{$sm << N$}. 
The purpose of this sample memory is to hold nodes' IDs that will be selected to create the output set $\sigma^{'}$ for current and future rounds.
When processing the input set $\sigma$, each encountered node ID $j$ is used to update the node's occurrence table $\Phi$ with its current occurrence count $\Phi_{j}$ (line 2).
If the sample memory $\Gamma$ is not already full and does not contain node $j$'s ID, it is added to the sample memory (line 4). 
If full, the node computes the probability of inserting node $j$ into the sample memory, denoted as $p_j$ (line 6), as the minimum known occurrence learned so far $min$ divided by the actual occurrence of node $j$, $\Phi_{j}$. 
The values $min$ and $\Phi_{j}$ are known from the occurrence table maintained in the tracking component.
If selected for insertion in the sample memory $\Gamma$, the ID of node $j$ replaces another existing entry selected uniformly at random.
Finally, an ID is selected uniformly at random from $\Gamma$ to fill the output set $\sigma^{'}$ (lines 10 and 11).

\begin{algorithm}
\caption{Local debiasing component for a node receiving the set of identifiers $\sigma$}
\KwIn{Input stream $\sigma$ in round $r$}
\KwOut{Output stream $\sigma^{'}$}
\SetKwInOut{Data}{Data}
\label{alg:Occstart}
\Data{Set of $sm$ identifiers $\Gamma$ \\
      }
      
\For{$j \in \sigma$}{
    $\Phi_{j} \leftarrow \Phi_{j} + 1$\;
    \If{$|\Gamma| < sm$}{
        $\Gamma \leftarrow \Gamma \cup \{j\}$\;
    }
    \Else{
        $min \leftarrow \min \Phi_{j}$;
        $p_j \leftarrow \frac{min}{\Phi_{j}}$\;
        \uIf{$\text{rand()} \leq p_j$}{
            choose unoformly at random $k_1$ from $\Gamma$\;
            $\Gamma \leftarrow (\Gamma \setminus \{k_1\}) \cup \{j\}$\;
        }
    }
        choose uniformly at random $k_2$ from $\Gamma$\;
        $\sigma^{'} \leftarrow \sigma^{'} \cup \{k_2\}$\;
}
\end{algorithm}

\subsection{Secret collaborative debiasing}
 \label{ss:ms}

We provide \AP's trusted nodes the capacity to recognize their trusted peers hidden in the mass. 
Further, we enhance their behaviors compared with honest nodes by allowing them to exchange and aggregate the information they have collected on the dissemination of identifiers.
In turn, the debiasing executed by trusted nodes is based on a larger amount of information, enabling a more accurate cleansing of the received sets of identifiers. 
Consequently trusted nodes build less biased views and serve as a source of more representative samples of the system's membership\bluestrike{.} \red{thereby benefiting the correct nodes.}



\subsubsection{Mutual authentication}
To allow trusted nodes to verify beforehand whether they are interacting with other trusted nodes, we rely on a secure mutual authentication protocol, executed by all nodes before sending push or pull messages to a chosen neighbor. 
We assume that each node possesses a symmetric secret key. 
In \AP, untrusted nodes generate a random secret key during the initialization phase. In contrast, trusted nodes share a common secret key that is provisioned during the remote attestation phase.


The mutual authentication protocol between two nodes  $A$ and  $B$ operates as follows. 
First, $A$ generates a pseudo-random number $r_A$ and sends it to $B$ via a cryptographic challenge. 
In turn, $B$ generates another pseudo-random number $r_B$ and computes the hash of the concatenation of $r_A$ and $r_B$,  $H(r_A \cdot r_B)$, and encrypts it with its own secret key obtaining $[H(r_A \cdot r_B)]_{K_{B}}$. 
Then, $B$ sends $r_B$ and $[H(r_A \cdot r_B)]_{K_{B}}$ to $A$.  
Upon reception, $A$ computes $H(r_A \cdot r_B)$ and deciphers $[H(r_A \cdot r_B)]_{K_{B}}$ using its own secret key $K_{A}$.
If the two values are identical (\ie, $A$ and $B$ share the same secret key), $A$ can identify $B$ as trusted.
Then, $A$ sends $[H(r_B \cdot r_A)]_{K_{A}}$ to $B$. 
Like $A$, $B$ deciphers this encrypted hash using its own secret key $K_B$ and compares it with $H(r_B \cdot r_A)$. 
If the two are equal, $B$ can also identify $A$ as trusted and the protocol's execution is successful.
Note that in this protocol, if one of the two considered nodes is untrusted, no information about the trustworthiness of the trusted node is revealed, keeping the identity of trusted nodes secret.


\subsubsection{Remembering trusted peers and aggregating tracking components}
\label{sec:merge}

We aim to have trusted nodes function as a group of actors who track the dissemination of identifiers they observe, collectively sharing this information to accelerate the detection of nodes attempting to over-represent themselves. 
Ideally, one could devise a collaborative tracking oracle where all trusted nodes know one another, sharing and using a single and global tracking component. 
Although not practically feasible, this oracle would allow for the instant tracking of identifier propagation within the system. 
Essentially, the oracle represents a configuration in which all trusted nodes merge their tracking components during each round.


With \AP, we aim to approximate this oracle by employing well-known gossip-based aggregation techniques~\cite{gossip_aggregation}. 
The key is to merge tracking components in a commutative and associative manner between subsets of trusted nodes in every round. 
Based on the work of Jelasity et al.~\cite{gossip_aggregation}, we select the average operation as the candidate for merging tracking components.
Eventually, with sufficient merge operations between trusted nodes, they will possess tracking components that expose the same information.
Whether using a tracking data structure instantiated as a key-value store or a count-min-sketch, the merge operation involves averaging each corresponding entry from both structures (each entry from the same key in the case of key-value stores, or each matrix entry for count-min-sketches).


Since trusted nodes play a crucial role in accelerating the debiasing operations of \AP, each trusted node maintains a \emph{trusted peer list} of the last $M$ trusted node identifiers it has contacted. This list is updated as the trusted node contacts other trusted peers by replacing the oldest entry with each newly acquired trusted node identifier. 
During every round in \AP, each trusted node contacts the $M$ peers from their trusted peer list, sending them its tracking component before receiving theirs. The trusted node then merges \bluestrike{the received tracking components with its own.}\red{each received tracking component with its own.}
To ensure that the identity of trusted nodes remains undetectable, honest nodes perform similar operations, \eg maintaining a random list of $M$ identifiers and contacting them every round, although they do not merge as they cannot successfully execute the mutual authentication protocol.

%% file: sections/eval.tex
\section{Evaluation}\label{sec:evaluation}

We compare the Byzantine-tolerance of \AP against its competitor \brahms, as well as another extension of~\brahms:~\basalt~\cite{basalt:2023}.
In a nutshell,~\basalt employs the min-wise independent permutation technique of~\brahms to build the complete view of each node.
Basically, each node picks $l_1$ different random seeds to define $l_1$ hash functions and uses the min-wise independent permutation technique to select node IDs that minimize the functions' output.
Once a \basalt node sees all the IDs in the system, its view will no longer change and is considered complete.
To add dynamicity to the network connectivity and continuously generate fresh views, nodes periodically refresh some of the hash function's seeds used in the min-wise independent permutation technique.
As described in the initial~\basalt paper, we consider the value of 1 reset seed per round as the refresh rate, implying that each seed is reset every $l_1$ round on average.

We used a simulation infrastructure implemented in Rust to conduct a simulation campaign and assess the extent to which~\AP,~\brahms and~\basalt provide resilience against Byzantine nodes attempting to manipulate the view of honest nodes by over-representing themselves.
Our simulation infrastructure is also based on the official implementation of~\basalt, available on GitHub~\cite{basalt_simu}.
To assess each protocol's resilience, we measure the average proportion of malicious identifiers present in the local views of non-Byzantine nodes at the end of the simulation runs.
We run simulations over a system composed of $N = 10,000$ nodes.
Views are composed of $l_1 = 160$ entries.
We set the parameters $\alpha$, $\beta$ and $\gamma$ to $1/3$, and $l_2 = 160$.
Byzantine nodes run the attack described in section~\ref{sec:sysmodel}.
Each simulated configuration executed over 200 rounds represents a system with $f \times N$ malicious nodes, $t \times N$ trusted nodes, and $(1-f-t) \times N$ honest nodes.
Once over, we collect the logs of each node and tear down the simulator.

\begin{figure}[!t]
  \centering
  \includegraphics[width=1\linewidth]{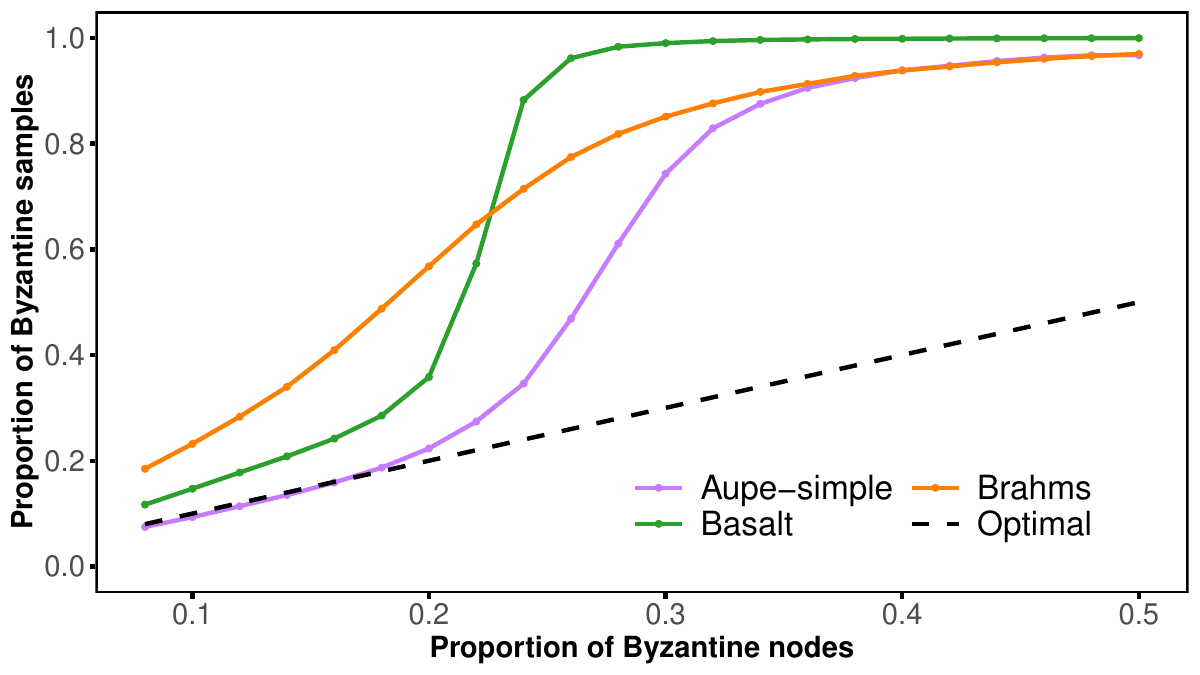}
  \caption{\AP's Byzantine resilience.}
  \label{fig:resilience_all}
\end{figure}

\red{Our evaluation of Aupe consider a key-value approach for the tracking component and a sample memory of size 100 for the debiasing component.}
To evaluate the effectiveness of \AP's debiasing strategy alone, we first consider a system with no trusted nodes.
We refer to this \AP variant as \emph{Aupe-simple}.
We vary the proportion $f$ of Byzantine nodes from $8\%$ to $50\%$ with steps of $2\%$, and compare \emph{Aupe-simple} to its two competitors.
Figure~\ref{fig:resilience_all} depicts the evolution of the average proportion of Byzantine identifiers in the views of non-Byzantine nodes for each protocol as $f$ increases.

As expected, \emph{Aupe-simple} always provides views composed of fewer Byzantine IDs than~\brahms.
While \brahms generates view polluted with 77\% of Byzantine IDs when $f = 26\%$, \emph{Aupe-simple} only reaches 46\%.
This is because \emph{Aupe-simple} identifies frequent IDs in the sets of pushed and pulled IDs and selects them less often than infrequent IDs for the construction of its push (\ie $\alpha \times l_1$) and pull (\ie $\beta \times l_1$) view subparts.
\emph{Aupe-simple} also surpasses~\basalt in all configurations and goes beyond the limit reached by~\basalt: when $f > 24\%$,~\basalt output views composed of more than 90\% of malicious IDs, while \emph{Aupe-simple} keeps the proportion of Byzantine IDs below, and only reaches about 80\% of malicious IDs in its views when $f > 32\%$.

\begin{figure}[!b]
  \centering
  \includegraphics[width=1\linewidth]{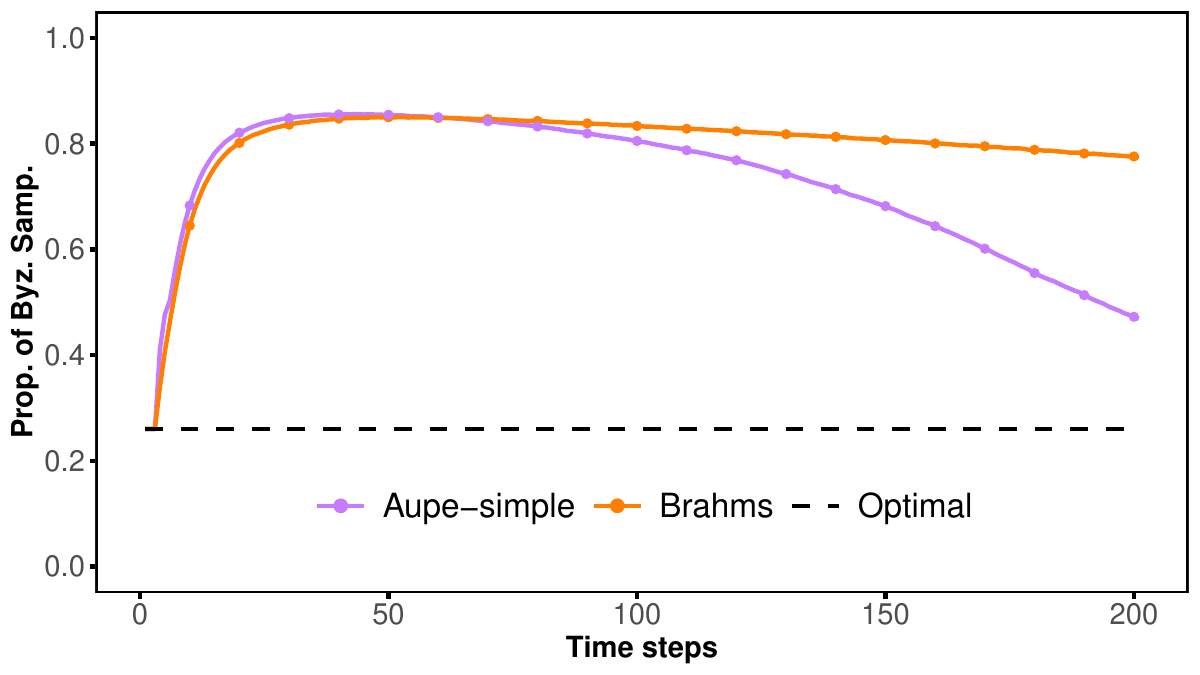}
  \caption{Evolution of proportion of Byzantine samples in the system, for $f=26\%$.}
  \label{fig:resilience_detail_view}
\end{figure}

\begin{figure}[!t]
  \centering
  \includegraphics[width=1\linewidth]{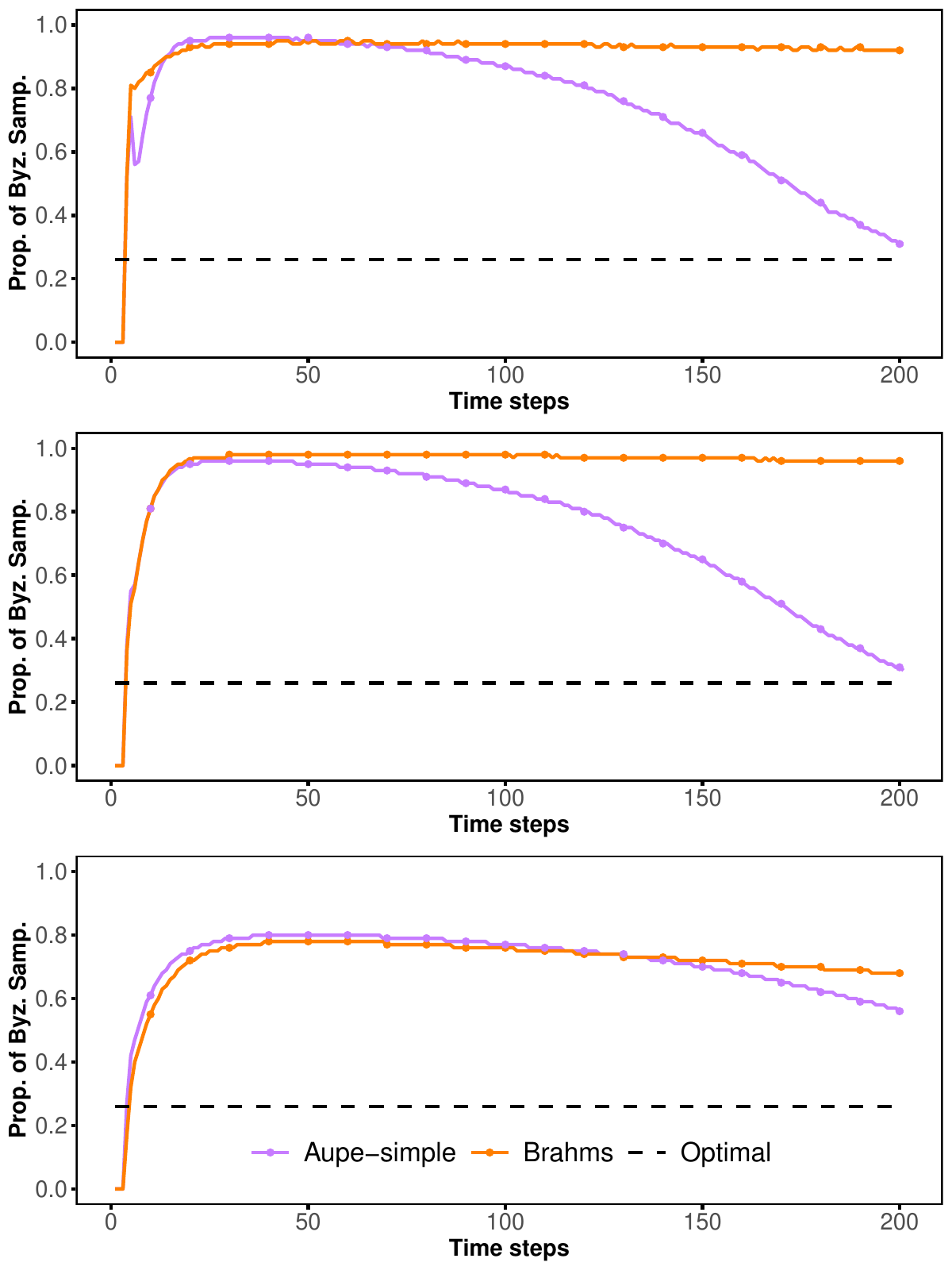}
  \caption{Evolution in time of Byzantine sample proportion inside the push view subpart (top figure), pull view subpart (middle figure) and history sample view subpart (bottom figure) for $f=26\%$.}
  \label{fig:resilience_detail_view_all}
\end{figure}

To better understand the impact of \AP's debiasing strategy, Figure~\ref{fig:resilience_detail_view} shows the evolution over time of the proportion of Byzantine IDs in the views of non-Byzantine nodes for $f = 26\%$.
In addition, Figure~\ref{fig:resilience_detail_view_all} shows the evolution of the proportion of Byzantine IDs in the push, pull, and history sample view subparts for $f = 26\%$.
At the end of this configuration's simulation, \emph{Aupe-simple} provides views composed with 46\% of malicious IDs whereas~\brahms reaches 77\%.
Detailed in Figure~\ref{fig:resilience_detail_view_all}, the push and pull view subparts of \emph{Aupe-simple} are significantly less polluted than those of~\brahms, 31\% and 30\% of Byzantine IDs, respectively, against 92\% and 96\% for~\brahms.
In turn, this also positively impacts the history sample view subpart, as \emph{Aupe-simple} nodes are allowed to discover new honest nodes at a faster pace than in~\brahms.

\subsection{On the impact of collaborative debiasing}

\begin{figure}[!t]
  \centering
  \includegraphics[width=1\linewidth]{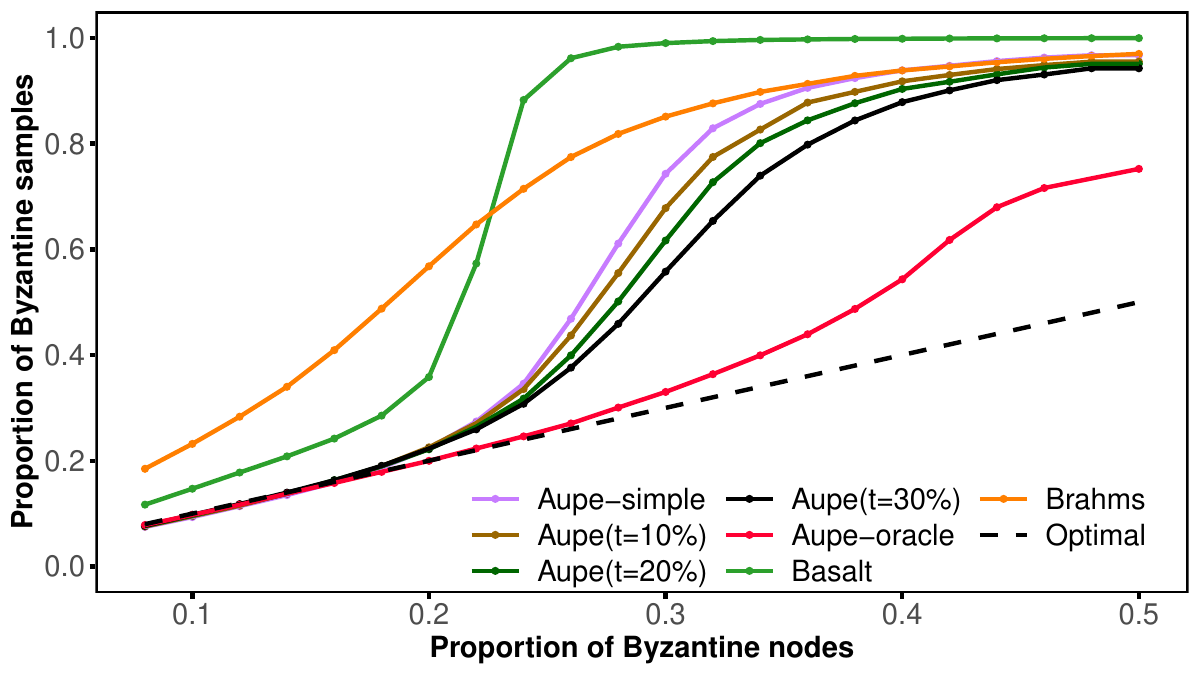}
  \caption{Impact of collaborative debiasing on \AP.}
  \label{fig:resf}
\end{figure}

To study the impact of introducing trusted nodes that cooperate to track the spread of IDs in the system, we first capture the potential of this collaborative approach by evaluating the oracle described in section~\ref{sec:merge}, referred to as \emph{Aupe-oracle} in the following figures.
Additionally, we examine three variants of \AP by introducing $t=10\%$, $t=20\%$ and $t=30\%$ of trusted nodes in the system, respectively.
The number of remembered trusted nodes, denoted $M$, is set to a low value of 10 in order to operate only a limited number of tracking component merges per round.
Figure~\ref{fig:resf} illustrates the evolution of the average proportion of Byzantine IDs in the views of non-Byzantine nodes for each protocol as $f$ increases.

\begin{figure}[!b]
  \centering
  \includegraphics[width=1\linewidth]{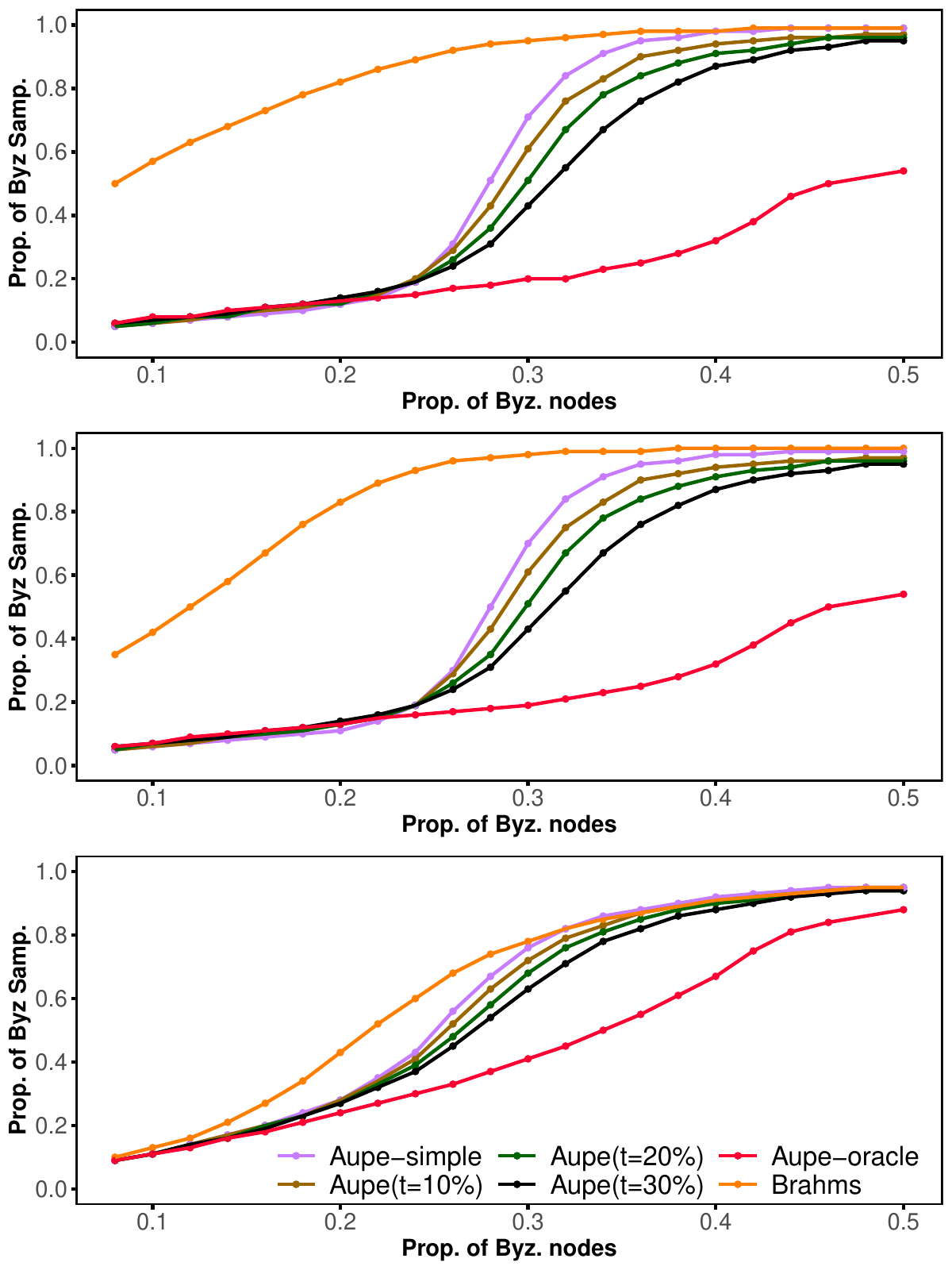}
  \caption{Byzantine sample proportion inside the push view subpart (top figure), pull view subpart (middle figure) and history sample view subpart (bottom figure) when varying f.}
  \label{fig:resilience_detail_f}
\end{figure}

Furthermore, Figure~\ref{fig:resilience_detail_f} reveals the impact of \AP's collaborative debiasing on each view subpart, \eg push, pull, and history sample.
Comparatively, \emph{Aupe-oracle} outperforms all other protocols, providing views with a proportion of Byzantine samples close to the optimal until $f$ exceeds 40\%.
However, while its push and pull view subparts are already optimal, the history sample view becomes polluted with Byzantine IDs as they propagate faster than honest and trusted IDs.

As the proportion of trusted nodes increases, \AP's variants become more resilient than \emph{Aupe-simple}.
This is due to the high effectiveness of the collaborative tracking strategy, as depicted in Figure~\ref{fig:resilience_detail_f}, the push and pull view subparts of each of \AP's variants become significantly less polluted when trusted nodes are added to the system.

\begin{figure}[!t]
  \centering
  \includegraphics[width=1\linewidth]{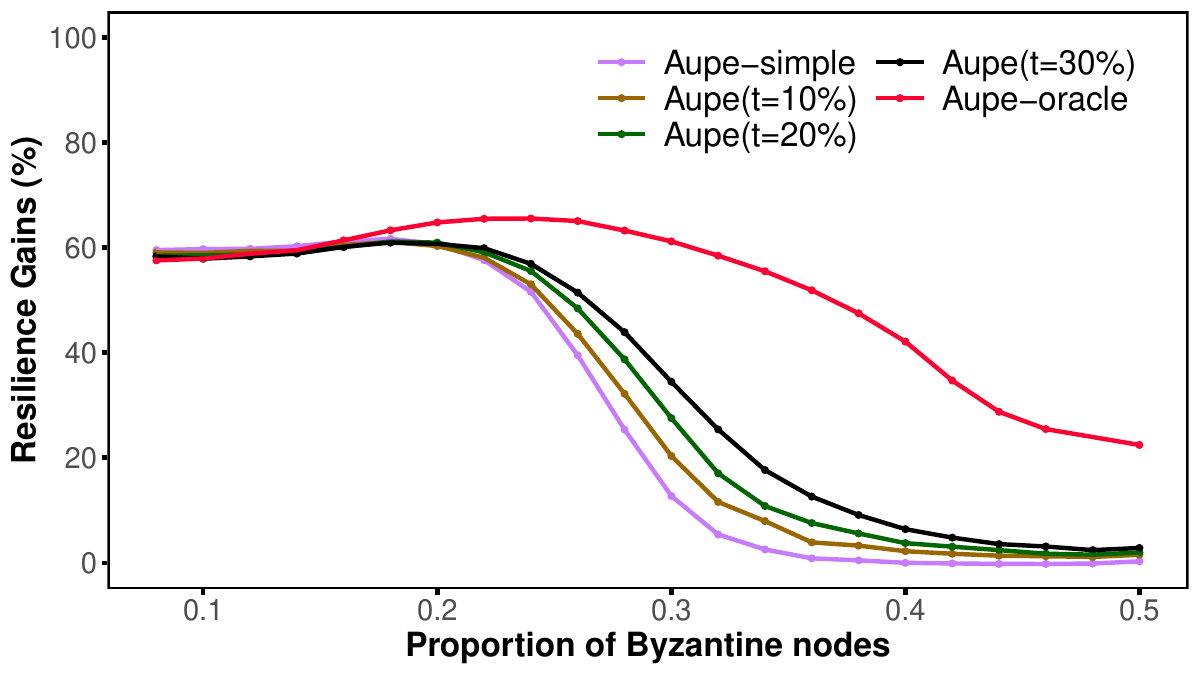}
  \caption{\AP's resilience gains compared to \brahms (the higher the better).}
  \label{fig:resGains}
\end{figure}

\begin{figure}[!b]
  \centering
  \includegraphics[width=1\linewidth]{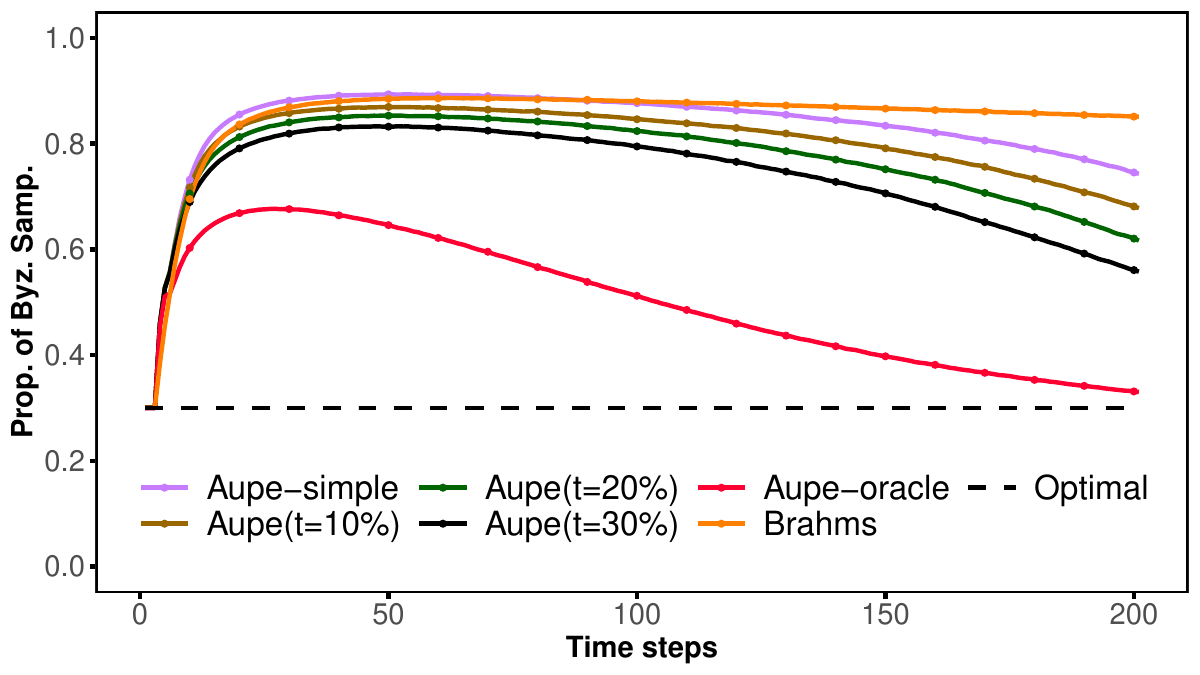}
  \caption{Evolution of merge for 200 rounds. $f=30\%$.}
  \label{fig:resilience_time_1}
\end{figure}

In Figure~\ref{fig:resGains}, the graph illustrates the resilience gains of \AP compared to \brahms.
This refers to the percentage reduction of Byzantine samples in the view of non-Byzantine nodes.
Overall, for values of $f < 24\% $, \AP provides up to 60\% resilience gains with regards to \brahms.
When f increases from $24\%$ to $40\%$, the attack led by the adversary become more effective, resulting in the resilience gains gradually dropping to 0\% when $f = 40\%$.
In these configurations, the specific role played by trusted nodes becomes evident.
The presence of more trusted nodes helps mitigate the reduction in resilience gains.
With $f=30\%$, incorporating 30\%, 20\% and 10\% of trusted nodes reduce the proportion of malicious IDs in \AP's views by 34\%, 27\% and 20\%, respectively, in comparison to \brahms.
Figure~\ref{fig:resilience_time_1} shows the evolution of the proportion of Byzantine IDs in the view of non-Byzantine nodes over time for $f = 30\%$.
While each of \AP's variants initially show similar behavior, introducing a greater part of trusted nodes helps them gather more information about the spread of Byzantine IDs over time.
This, in turn, enables them to reduce the proportion of Byzantine samples in their views and consequently, in the views of honest nodes.

%% file: sections/sota.tex
\section{Related Work}\label{sec:related}

Peer sampling protocols have been mostly studied in non-adversarial environments. Cyclon~\cite{voulgarisCYCLON2005}, Newscast~\cite{tolgyesi2009adaptive}, and the more generic protocol framework for gossip-based peer sampling of Jelasity et al.~\cite{jelasityGossipbased2007} efficiently handle churn and quickly discover other nodes in the network.

A few studies targeted protocols tolerant to Byzantine behaviors~\cite{Bortnikov:2008, Anceaume:2013, basalt:2023, antonov2023securecyclon, Pigaglio:2022, JESI20102086}.
In \emph{SPS}~\cite{JESI20102086}, nodes employ a fault detection mechanism to identify and blacklist nodes acting maliciously in the system.
This protocol remains, however, vulnerable to rapid flooding attack as correct nodes cannot identify and blacklist attackers before being overwhelmed by them and isolated.
SecureCyclon~\cite{antonov2023securecyclon} further investigates fault detection mechanisms to prevent additional protocol violations.

Instead of relying on detection mechanisms, the seminal work \brahms~\cite{Bortnikov:2008} employs a min-wise permutation technique~\cite{broder1998min} to ensure that a subpart of the view is composed of an actual uniform sample of the set of identifiers each node has seen.
Additional countermeasures to specific Byzantine attacks complement the approach to prevent node isolation and eviction.
Similarly to \brahms, \basalt~\cite{basalt:2023} exploits a greedy epidemic procedure towards random nodes that are implicitly defined using min-wise independent permutations.
In \basalt, each node composes its local view from the current state of this greedy epidemic procedure, extending the use of \brahms's min-wise independent permutation techniques to build the entirety of the view instead of only a subpart.
The authors of \raptee~\cite{Pigaglio:2022} introduced trusted nodes operating on trusted execution environments to improve the resilience of \brahms.
Communications between the different types of nodes (honest-to-honest, honest-to-trusted, and trusted-to-trusted) are designed to accelerate the spread of identifiers among trusted nodes and slow down their propagation from honest nodes (possessing slightly more polluted views) to trusted nodes.
As a result, \raptee improves the resilience of \brahms by up to 21\% when considering 10\% of trusted nodes and 10\% of Byzantine nodes in the system.
Similar to our work, trusted nodes in \raptee can act as a source of slightly less biased views of honest nodes.
In our work, we go one step further by enabling collaboration between trusted nodes, allowing them to share and collectively make use of the knowledge they obtained on the actual dissemination of identifiers.
Doing so, \AP outperforms \raptee by providing up to 60\% of resilience gains for $f \le 24\%$ while mitigating the impact of the adversary's attack, even when having control over 40\% of the nodes.



%% file: sections/conclusion.tex
\section{Conclusion}\label{sec:conclusion}

We introduced \AP, the first Byzantine-tolerant random peer sampling protocol that utilizes collaborative trusted debiasing. The inclusion of trusted nodes in the system model enabled the collaborative tracking of identifier spread within the system and the local debiasing of Byzantine nodes in the set of received descriptors. Through simulations involving 10,000 nodes, \AP demonstrated superior resilience compared to state-of-the-art solutions. It exhibited near-perfect resilience, even in the presence of an adversary controlling 26\% of the nodes. By incorporating as few as 10\% trusted nodes, \AP increased the resilience of ~\brahms by up to 60\%, while mitigating the impact of the adversary's attack, even under 40\% faulty nodes.


